\documentclass[reprint,superscriptaddress,amssymb,amsmath,aps,showpacs,10pt,longbibliography,prb]{revtex4-2}
\usepackage{graphicx}
\graphicspath{{figures/}}
\usepackage{color}
\usepackage[colorlinks,bookmarks=false,citecolor=darkblue,linkcolor=red,urlcolor=blue]{hyperref}

\definecolor{darkred}{rgb}{0.7,0.0,0.0}
\definecolor{darkblue}{rgb}{0,0.02,0.45}

\definecolor{darkgreen}{rgb}{0.02,0.45,0.0}
\definecolor{hlcolor}{RGB}{10,186,181}

\begin{document}

%------------------------------------------------------------------------------------------------

\title{Tunable Dirac nodal line in orthorhombic RuO$_2$}

\author{Ece Uykur}
\email{e.uykur@hzdr.de}
\affiliation{Helmholtz-Zentrum Dresden-Rossendorf, Inst Ion Beam Phys \& Mat Res, D-01328 Dresden, Germany}

\author{Oleg Janson}
\email{o.janson@ifw-dresden.de}
\affiliation{Institute for Theoretical Solid State Physics, Leibniz Institute for Solid State and Materials Research Dresden, 01069 Dresden, Germany}

\author{Victoria A. Ginga}
\affiliation{Felix Bloch Institute for Solid-State Physics, University of Leipzig, 04103 Leipzig, Germany}

\author{Marcus Schmidt}
\affiliation{Max Planck Institute for Chemical Physics of Solids, 01067 Dresden, Germany}

\author{Nico Giordano}
\affiliation{Deutsches Elektronen-Synchrotron DESY, 22607 Hamburg, Germany}

\author{Alexander A. Tsirlin}
\email{altsirlin@gmail.com}
\affiliation{Felix Bloch Institute for Solid-State Physics, University of Leipzig, 04103 Leipzig, Germany}

%\date{\today}

\begin{abstract}
Pressure evolution of RuO$_2$ is studied using single-crystal x-ray diffraction in a diamond anvil cell, combined with \textit{ab initio} band-structure calculations. The tetragonal rutile structure transforms into the orthorhombic CaCl$_2$-type structure above 13\,GPa under quasi-hydrostatic pressure conditions. This second-order transition is ferroelastic in nature and accompanied by tilts of the RuO$_6$ octahedra. Orthorhombic RuO$_2$ is expected to be paramagnetic metal, similar to ambient-pressure RuO$_2$. It shows the increased $t_{2g}-e_g$ crystal-field splitting that is responsible for the pressure-induced color change. It further features the Dirac nodal line that shifts across the Fermi level upon compression. 
\end{abstract}

\maketitle

%--------------------------------------------------------------------------------------------------

\section{Introduction}
Ruthenium dioxide belongs to the family of rutile-type compounds. It is a good metal that boasts widespread applications in thermometry~\cite{li1986} and catalysis~\cite{over2012}. Recently, it received renewed interest in the context of altermagnetism, the state with zero net magnetization and lifted Kramers degeneracy~\cite{smejkal2022}. Signatures of altermagnetism have been mostly reported in RuO$_2$ thin films~\cite{feng2022,bai2023,fedchenko2024,wang2024}, whereas bulk samples show conventional paramagnetic behavior of a Fermi liquid~\cite{wenzel2024} and evade magnetic order according to muon spectroscopy~\cite{hiraishi2024} and neutron diffraction~\cite{kessler2024,kiefer2024}. The electronic structure of bulk RuO$_2$ probed by optical~\cite{wenzel2024} and photoemission~\cite{liu2024} spectroscopies is well described by spin-unpolarized band-structure calculations and does not show the spin splitting expected in altermagnets. However, it is plausible that RuO$_2$ lies in the vicinity of unconventional states, not only magnetic but also superconducting, in view of superconductivity reported in strained thin films~\cite{uchida2020,ruf2021}. Moreover, a Dirac nodal line lying only 45\,meV below the Fermi level has been reported in the band structure of bulk RuO$_2$~\cite{jovic2018,wenzel2024} and could be responsible~\cite{sun2017} for the promising spin transport properties of this material~\cite{bai2022,karube2022,bose2022}. 

External pressure transforms the ambient-pressure (AP) tetragonal polymorph into the orthorhombic \mbox{(HP-I)} polymorph that manifests as a distorted version of the rutile structure~\cite{haines1993}, also known as the CaCl$_2$-type structure after a similar transformation that takes place in CaCl$_2$ on cooling~\cite{howard2005}. This orthorhombic HP-I phase is a possible playground for tailoring RuO$_2$ properties because it is similar to the parent rutile structure, yet it involves additional degrees of freedom related to tilts of the RuO$_6$ octahedra. Experimental studies suggest that RuO$_2$ changes color from black to orange and eventually to yellow under pressure~\cite{rosenblum1997,white2024}. It also loses its metallicity above 28\,GPa~\cite{white2024}. The interpretation of these phenomena is hindered by the absence of accurate structural data under pressure. The crystal structure of the HP-I polymorph was refined only at one pressure using neutron diffraction~\cite{haines1997}. X-ray diffraction data were collected under nonhydrostatic conditions~\cite{haines1993,armstrong2020} that facilitate coexistence of the HP-I phase with higher-pressure polymorphs, thus further complicating the analysis.

In the following, we report the full structure determination for the AP and HP-I polymorphs of RuO$_2$ probed under quasi-hydrostatic conditions at pressures up to 37\,GPa. Using experimental lattice parameters and atomic positions we calculate band structures of both polymorphs and identify these phases as robust paramagnetic metals. The color change is attributed to the increased crystal-field splitting that shifts the minimum of the reflectivity into the visible range. Although HP-I RuO$_2$ does not become altermagnetic, it shows an interesting evolution of the Dirac nodal line across the Fermi level. This evolution is driven by the orthorhombic deformation of the rutile structure and suggests that strain tuning may be instrumental in modifying electronic properties of RuO$_2$.

%-------------------------------

\section{Methods}

Single crystals of RuO$_2$ were prepared by chemical vapor transport as described previously~\cite{wenzel2024}. A single crystal with the dimensions of $30\times 20\times 10$\,$\mu$m$^3$ was loaded into a diamond anvil cell (DAC) with a 300\,$\mu$m culet diameter and filled with neon gas as pressure-transmitting medium. Pressure was determined by ruby luminescence~\cite{mao1986}. Neon ensures quasi-hydrostatic conditions up to 15\,GPa~\cite{klotz2009} and develops relatively weak non-hydrostaticity at higher pressures, as confirmed by the sharp ruby peaks over the pressure range of our experiment.

Single-crystal x-ray diffraction (XRD) data were collected at room temperature on the P02.2 beamline~\cite{p02-2} of the PETRA III synchrotron (DESY, Hamburg, Germany) using x-rays with the wavelength of 0.2910\,\r A and Perkin Elmer XRD 1621 detector. Diffraction images were collected by a $\varphi$-rotation of the pressure cell between $-30$ and $+30^{\circ}$ with the step of $0.5^{\circ}$ and integrated using \texttt{CrysAlisPro} software~\cite{crysalispro}. Structure refinements were performed in Jana2006~\cite{jana2006}.

The electronic structure of RuO$_2$ at different pressures was obtained within the
framework of density functional theory (DFT) using the
\texttt{Wien2K}~\cite{wien2k,blaha2020} and \texttt{FPLO}~\cite{fplo} codes
with the Perdew-Burke-Ernzerhof (PBE) exchange-correlation
potential~\cite{pbe96} and experimental structural parameters determined by
single-crystal XRD. To account for the spin-orbit (SO) coupling, all
calculations were performed in the full relativistic mode. FPLO was further
used to calculate Wannier projections~\cite{FPLO-WF} of PBE-SO bands onto i)
all $4d$ orbitals of Ru and ii) the $t_{2g}$ manifold thereof.  The former were
used to construct the local Hamiltonian $H_0$ whose eigenvalues
$\varepsilon_{1..5}$ (for ease of notation, we take only one eigenvalue of each
Kramers' doublet) allowed us to estimate the crystal-field splitting as
$\Delta_{t_{2g}-e_g}=\frac12\left(\varepsilon_4 + \varepsilon_5\right) -
\frac13\left(\varepsilon_1 + \varepsilon_2 + \varepsilon_3\right)$. The more compact
Hamiltonians based on $t_{2g}$ projections were used to visualize how the
energies and momenta of the nodal line evolve with pressure.

Additionally, crystal structures of different RuO$_2$ polymorphs were relaxed
at several constant volumes to obtain the equation of state. These calculations
were performed in \texttt{VASP}~\cite{vasp1,vasp2} using the PBE functional as
well as the strongly constrained and appropriately normed semi-local
SCAN~\cite{scan} functional that better reproduces the equilibrium volume of
RuO$_2$. 

%-------------------------------

\section{Results}

\subsection{Structural phase transition}

\begin{figure}[h!]
  \centering
  \includegraphics[width=0.5\textwidth]{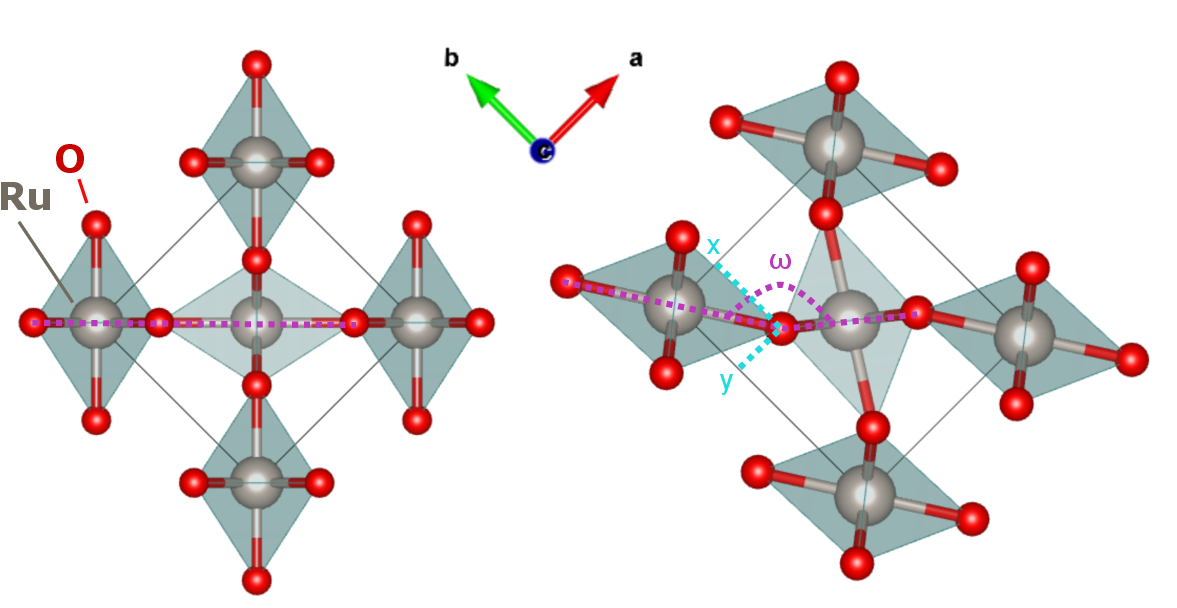}
    \caption{Low-pressure tetragonal (AP, left panel) and high-pressure orthorhombic (HP-I, right panel) polymorphs of RuO$_2$. The tilt angle $\omega$ and the fractional coordinates of oxygen ($x$, $y$) are labeled. The structures are visualized using \texttt{VESTA}~\cite{VESTA}.}
    \label{F1}
\end{figure}

\begin{figure}[h!]
  \centering
  \includegraphics[width=0.5\textwidth]{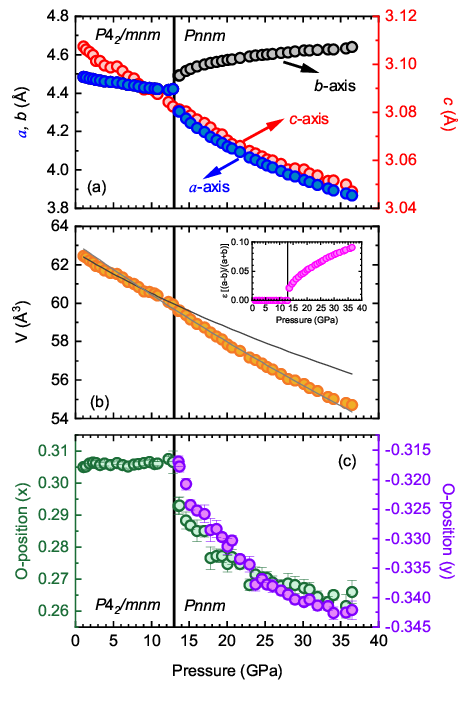}
    \caption{(a) Lattice parameters of RuO$_2$ as a function of pressure. (b) Pressure dependence of the unit-cell volume with a small kink clearly visible at the AP-to-HP-I transition around 13\,GPa. The solid gray lines are the fits to the equation of state as described in the text. The inset shows the orthorhombic strain, $\epsilon$. (c) Pressure evolution of the oxygen $x$- and $y$-coordinates in the AP and HP-I polymorphs. Note that symmetry requires $x=y$ in the tetragonal phase. }
    \label{F2}
\end{figure}

Our single-crystal XRD data (Fig.~\ref{F2}) reveal a gradual compression of the tetragonal structure (AP polymorph) up to 12\,GPa. The position of oxygen remains constant, similar to other tetragonal rutile-type oxides~\cite{hazen1981}, and only the lattice parameters change with pressure. Above 13\,GPa, the indexing of the reflections clearly indicates the symmetry change to orthorhombic with the space group $Pnnm$ that has been reported for the HP-I polymorph of RuO$_2$~\cite{haines1993}. The orthorhombic strain gradually increases with pressure and reaches $\epsilon=(a-b)/(a+b)=0.09$ at 36.5\,GPa. The pressure dependence of the $c$ parameter does not change slope at the transition, whereas the unit-cell volume shows a small kink but evolves continuously, indicative of a second-order phase transition.

Structure refinement for the HP-I polymorph reveals that the deformation (weak axial compression) of the RuO$_6$ octahedra remains almost constant across the pressure range of our study. However, the orthorhombic strain causes tilting of the octahedra, with the tilt angle $\omega$ serving as an order parameter of the transition. Its pressure evolution is typical of a second-order phase transition and follows 
\begin{equation}
 \omega(P)\sim[(P/P_c)-1]^{\beta}
\end{equation}
with $P_c=12.8$\,GPa and $\beta=0.35$. The tilt angle scales linearly with the orthorhombic strain, similar to CaCl$_2$ and related compounds~\cite{howard2005}. This linear scaling supports the scenario of a ferroelastic phase transition driven by a soft mode due to the octahedral tilting. 

\begin{figure}
  \centering
  \includegraphics[width=0.45\textwidth]{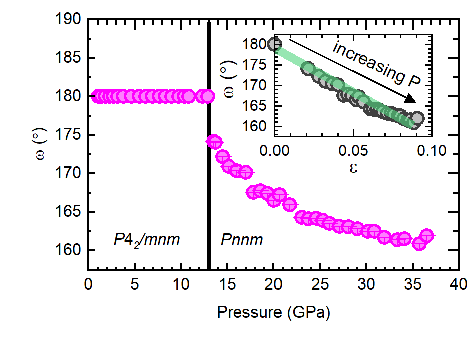}
    \caption{Tilt angle $\omega$ (Fig.~\ref{F1}, right) as a function of pressure. The inset shows the linear scaling of the tilt angle with the orthorhombic strain. }
    \label{F3}
\end{figure}

Previous studies reported different critical pressures of the AP-to-HP-I transition in RuO$_2$ depending on the pressure medium: 5.0\,GPa (silicon oil~\cite{haines1993}), 7.6\,GPa (sodium chloride~\cite{ono2011}), and 11.8\,GPa (argon~\cite{rosenblum1997}), all values taken at room temperature. It appears that the transition pressure increases as deviations from hydrostaticity decrease~\cite{klotz2009}, so it seems natural that our experiments performed in quasi-hydrostatic conditions return an even higher value of about 13\,GPa. 

We note in passing that the reflections of the HP-I polymorph could be traced in our data up to 36.5\,GPa, and no other reflections, except those from the diamond anvil cell, have been observed in this pressure range. At 37.8\,GPa, the reflections of the HP-I polymorph abruptly disappear, and a cubic phase reported as the second high-pressure polymorph of RuO$_2$~\cite{haines1993,haines1996,white2024} was observed. A detailed structure of this cubic phase has not been resolved in the present experiment and will be an interesting topic for future investigation. 

%-------------------------------

\subsection{Equation of state}

\begin{figure}
  \centering
  \includegraphics[width=0.45\textwidth]{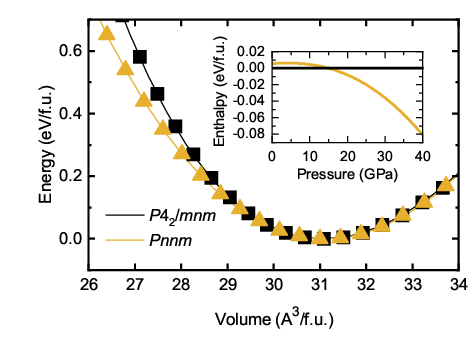}
    \caption{Energies of the AP and HP-I polymorphs as a function of unit-cell volume. The inset shows the calculated pressure-dependent enthaply of the HP-I polymorph relative to the enthalpy of the AP polymorph taken as zero for reference.}
    \label{F4}
\end{figure}

Fig.~\ref{F4} shows energies of the AP and HP-I polymorphs of RuO$_2$ at several fixed volumes. They merge around the equilibrium volume, but start to deviate on compression, with the HP-I polymorph showing a lower energy at a given volume and, therefore, the higher compressibility compared to the AP-polymorph. The same trend is seen experimentally; the decrease in the volume becomes faster above the transition. It corresponds to the additional volume reduction upon the tilting of the RuO$_6$ octahedra. This additional mechanism renders the HP-I structure more compressible compared to the AP structure.

\begin{figure}
  \centering
  \includegraphics[width=0.5\textwidth]{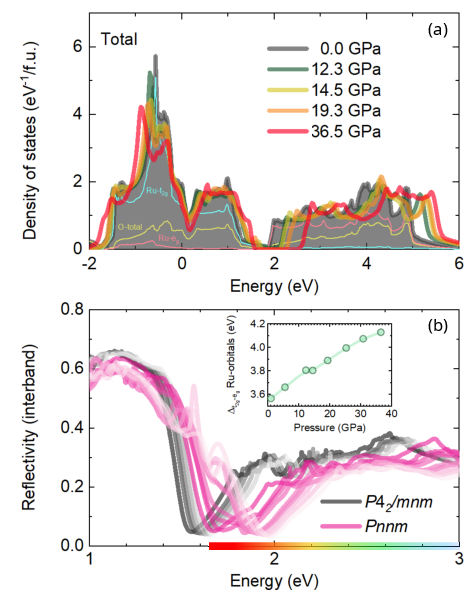}
    \caption{(a) Density of states as a function of pressure across the
structural phase transition. The contributions of Ru $t_{2g}$, Ru $e_g$ and O
are shown for the 0\,GPa spectrum and do not change significantly with pressure. (b) Computed
reflectivity (interband contribution) at different pressures. The gray and pink
traces show the tetragonal and orthorhombic phases, respectively; the color
opacity decreases with pressure. The reflectivity minimum shows a gradual blue
shift and above the structural phase transition falls into the visible energy
range shown by the color bar on the energy axis. Note that the energy axis has
been divided by 1.35 according to the band energy renormalization determined
from the ambient-pressure optical study~\cite{wenzel2024}. Inset: The crystal
field splitting $\Delta_{t_{2g}-e_g}$ determined from local Hamiltonians based
on Wannier projections onto Ru $4d$ states.}
    \label{F6}
\end{figure}
\begin{figure*}
  \centering
  \includegraphics[width=1\textwidth]{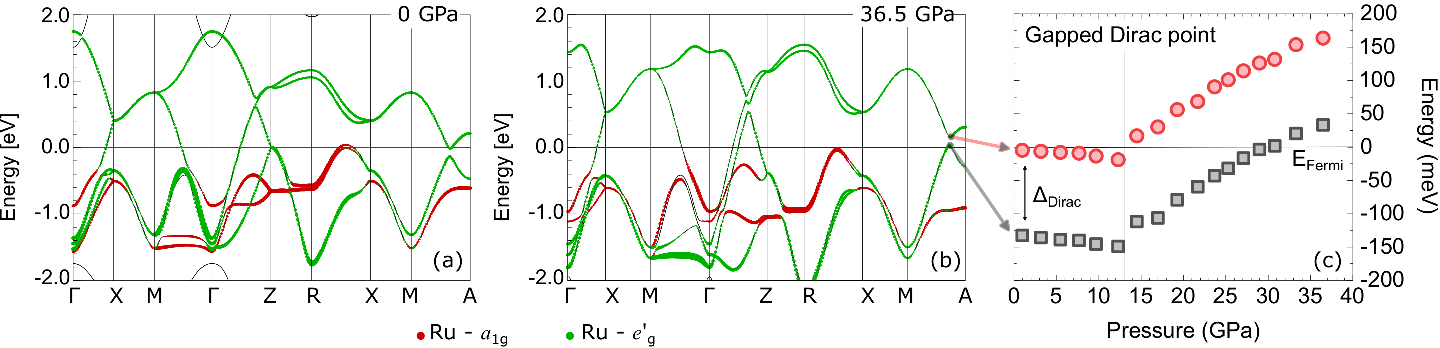}
    \caption{(a, b) Band structures of RuO$_2$ at 0\,GPa (tetragonal AP polymorph) and 36.5\,GPa (orthorhombic HP-I polymorph). The colored dots in the band dispersions show the contributions of the $a_{1g}$ and $e_g'$ orbitals of the Ru $t_{2g}$ manifold. The $a_{1g}$ bands remain almost fully filled under pressure. (c) Positions of the band minima and maxima along $M-A$ indicate the Dirac point that gradually shifts toward higher energies above the structural phase transition, whereas the Dirac gap remains almost unchanged. }
    \label{F5}
\end{figure*}

Fits of the energy-vs-volume curves with the Murnaghan equation of state return the equilibrium volume $V_0=62.141(6)$\,\r A$^3$/f.u., bulk modulus $B_0=287(1)$\,GPa, and pressure derivative of the bulk modulus $B_0'=4.41(8)$ in the AP-polymorph. The HP-I polymorph is described by $V_0=62.24(1)$\,\r A$^3$/f.u., $B_0=263(1)$\,GPa, and $B_0'=1.0(1)$. This polymorph is indeed more compressible because its bulk modulus only weakly increases with pressure. The calculated enthalpies indicate the AP-to-HP-I transition around 15\,GPa in good agreement with the experimentally determined transition pressure of 13\,GPa. All of the aforementioned results were obtained with the SCAN functional that underestimates the equilibrium volume of RuO$_2$ by 0.8\%. Using PBE leads to a more drastic mismatch with the 2.1\% overestimation of the equilibrium volume in agreement with the previous studies~\cite{tse2000}. The resulting transition pressure of 16.5\,GPa is, nevertheless, quite similar to the value obtained using the SCAN functional.

%-------------------------------

\subsection{Electronic structure}

Both optical~\cite{wenzel2024} and angle-resolved photoemission spectroscopy~\cite{liu2024} studies at ambient pressure consistently show that the electronic structure of bulk RuO$_2$ is well described by nonmagnetic calculations performed on the PBE+SO level without including correlation effects. We use the same computational method to assess possible magnetism of the HP-I polymorph. Calculations performed with the initial spin polarization converged to the nonmagnetic solution for all pressures studied in this work. Therefore, we conclude that orthorhombic RuO$_2$ should be a paramagnetic metal, similar to the AP polymorph. 

In RuO$_2$, the states near the Fermi level are formed by Ru $t_{2g}$ orbitals. The respective bands are fully separated from the higher-lying $e_g$ bands because of the large crystal-field splitting $\Delta_{t_{2g}-e_g}$. Our calculations show that this splitting increases with pressure, whereas the widths of the $t_{2g}$ and $e_g$ bands change only marginally (Fig.~\ref{F6}a). Moreover, the atomic SO coupling parameter $\zeta$ of Ru $4d$, which can be also estimated from the local Hamiltonian~\cite{kuzian21}, remains practically constant ($\zeta=115$\,meV at 5.5 GPa and 113\,meV at 36.5 GPa). The crystal-field splitting leads to a pronounced minimum in the optical conductivity of RuO$_2$. This minimum shifts toward higher energies under pressure as $\Delta_{t_{2g}-e_g}$ increases (Fig.~\ref{F6}b). At ambient pressure, the minimum lies in the near-infrared range, thus explaining the black color of RuO$_2$ crystals. Our calculations also explain the experimentally observed color change toward red at 13\,GPa and eventually toward yellow at higher pressures~\cite{white2024}. The structural phase transition accompanies the shift of the reflectivity minimum and the concomitant color change, but it does not trigger this change. In fact, Ref.~\cite{rosenblum1997} reports the appearance of the red color of RuO$_2$ already at 4.5\,GPa, well before the structural phase transition that was detected around 12\,GPa in the same experiment.

Although the density of states at the Fermi level decreases with pressure, orthorhombic RuO$_2$ remains a robust metal (Fig.~\ref{F6}a). The loss of the metallicity detected experimentally above 28\,GPa~\cite{white2024} must be related to other high-pressure phases that could be induced by non-hydrostatic effects and did not occur in our experiment performed under quasi-hydrostatic conditions.

\begin{figure*}[tb]
  \centering
  \includegraphics[width=1\textwidth]{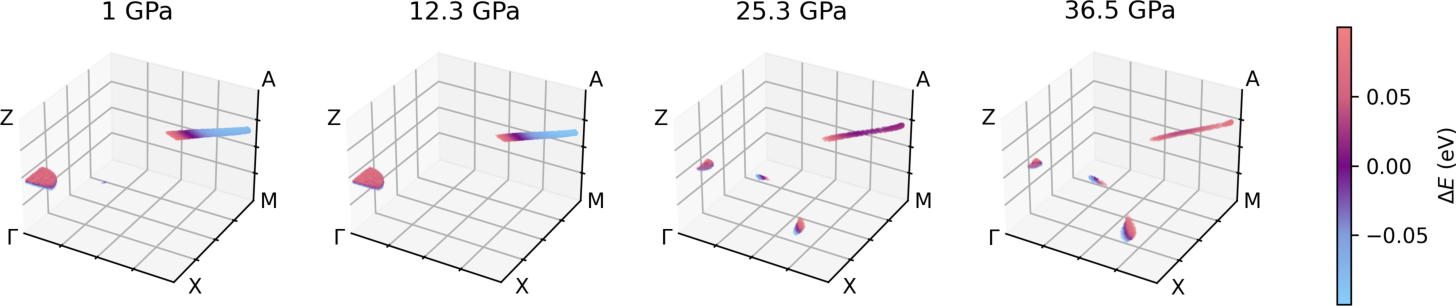}
    \caption{Visualization of the Dirac nodal line at four selected pressures: 1 and 12.3\,GPa correspond to the AP phase, 25.3 and 36.5\,GPa -- to the HP-I phase. The colored regions correspond to locations in the Brillouin zone (only the positive octant is shown) where the gap between two bands forming the Dirac point in Fig.~\ref{F5} is smaller than 100\,meV; the color coding indicates the difference $\Delta{}E = E(\vec{k}) - E(\text{Dirac})$ between the midpoint of this gap at the given momentum $E(\vec{k})$ and the corresponding value $E(\text{Dirac})$ at the gapped Dirac point along the M-A line. Only points with $|\Delta{}E|\leq100$\,meV are shown. The plotting was made using matplotlib~\cite{matplotlib}. }
    \label{F7}
\end{figure*}

At ambient pressure, RuO$_2$ features a Dirac nodal line lying 45\,meV below the Fermi level~\cite{wenzel2024}. This nodal line follows the diagonal in the $k_x-k_y$ plane with no dispersion along $k_z$ (Fig.~\ref{F7}). Our calculations suggest that this feature remains almost unchanged across the AP phase (Fig.~\ref{F7}). As the orthorhombic HP-I phase sets in, the Dirac crossing (gapped by the SO coupling) shifts toward higher energies and crosses the Fermi level around 20\,GPa (Fig.~\ref{F5}). The same plot reveals that the $\Delta_\text{Dirac}$ gap itself stays almost intact, suggesting that the nodal line would behave as a rigid object in momentum space, and would only change its energy position. The actual calculations reveal that this scenario is oversimplified: while the nodal line is clearly visible in the HP-I phase, it becomes ``thinner'' in the momentum space and significantly sharper along the energy axis (Fig.~\ref{F7}). These results suggest a very robust nature of the Dirac nodal line that persists despite the sizable orthorhombic strain setting on in the AP-I phase. Therefore, strain applied in the $ab$ plane of the RuO$_2$ structure can be an effective tuning parameter for adjusting the position of the Dirac nodal line in RuO$_2$. 

%-------------------------------

\section{Discussion and Summary}

The orthorhombic HP-I polymorph of RuO$_2$ forms upon the pressure-induced second-order ferroelastic phase transition. Our data obtained under quasi-hydrostatic conditions reveal this transition around 13\,GPa is in good agreement with the \textit{ab initio} theory. Experiments performed under non-hydrostatic conditions reported much lower values of 5 to 12\,GPa for the critical pressure~\cite{haines1993,rosenblum1997,ono2011}. This unusual sensitivity of RuO$_2$ to the pressure environment probably mirrors the variety of physical regimes observed in RuO$_2$ thin films~\cite{feng2022,uchida2020,ruf2021} and suggests that properties of this material can be tailored using strain. 

Although 13\,GPa marks the lower boundary of the thermodynamic stability of HP-I RuO$_2$, its upper boundary is more difficult to determine. Although we observed the HP-I polymorph up to 36.5\,GPa followed by the HP-II cubic polymorph at higher pressures, it is important to note that, in contrast to the AP-to-HP-I transition, the transformation from HP-I to HP-II involves a major rearrangement of the crystal structure~\cite{haines1993} and may be kinetically hindered. 

Orthorhombic RuO$_2$ is expected to be metallic and nonmagnetic. It retains the main features of the tetragonal phase, most notably, the Dirac nodal line in the vicinity of the Fermi level. Orthorhombic strain shifts the nodal line toward higher energies, such that it crosses the Fermi level around 20\,GPa. The nodal line becomes even sharper in energy, which is beneficial for its experimental observation. Overall, the straining in our experiments appears to be different from the strain in RuO$_2$ thin films that are typically grown on the $(110)$ surface of TiO$_2$ and thus offers an additional dimension for manipulating the electronic structure of RuO$_2$.

In summary, we explored the high-pressure orthorhombic (HP-I) polymorph of RuO$_2$ that forms upon the second-order ferroelastic transition around 13\,GPa. Band-structure calculations performed using experimental structural parameters ascribe the pressure-induced color change to the gradual increase in the $t_{2g}-e_g$ crystal-field splitting that shifts the minimum in the reflectivity of RuO$_2$ toward higher energies. 
The HP-I polymorph does not show the proclivity for magnetism and remains a robust metal in our calculations. The orthorhombic strain serves as a tuning knob for the band structure and especially for the Dirac nodal line that changes its energy with pressure and crosses the Fermi level around 20\,GPa.

\acknowledgments
We thank Vicky Hasse for her technical assistance with the crystal growth and Sahana R\"o{\ss}ler for characterizing the RuO$_2$ crystal used in this study. We acknowledge DESY (Hamburg, Germany), a member of the Helmholtz Association HGF, for the provision of experimental facilities. Parts of this research were carried out at the P02.2 beamline of PETRA III. Beamtime was allocated for the proposal I-20230876. The work in HZDR is funded by the DFG UY 63/2-1. O.J. was supported by the Deutsche Forschungsgemeinschaft (DFG, German Research Foundation) Project No.\ 247310070, and thanks U.\ Nitzsche for technical assistance.  The work in Leipzig was funded by the Deutsche Forschungsgemeinschaft (DFG, German Research Foundation) -- TRR 360 -- 492547816 (subproject B1).

\newpage

\section*{Appendix}

\begin{figure}[h!]
  \centering
  \includegraphics[width=0.4\textwidth]{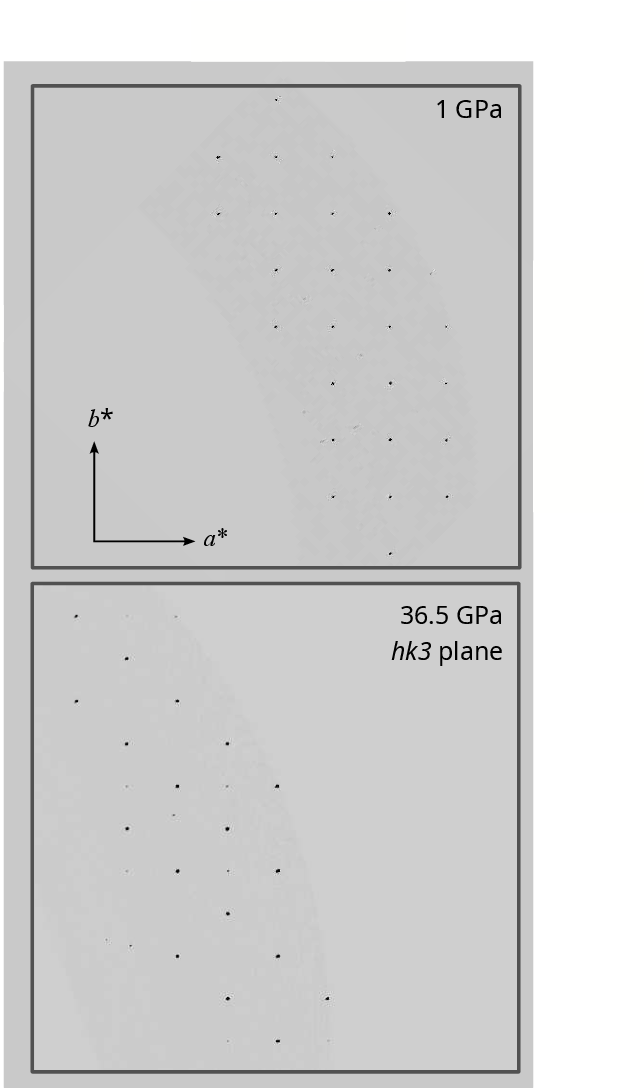}
    \caption{Reciprocal-lattice reconstruction in the $hk3$ plane for the AP phase at 1\,GPa (top) and HP-I phase at 36.5\,GPa (bottom) show the symmetry reduction from tetragonal to orthorhombic. The spurious reflections visible in the images arise from the diamond anvil cell. }
    \label{F8}
\end{figure}

\begin{table}[h!]
\centering
\caption{Details of the data collection and refined structural parameters for AP RuO$_2$ at 1.0 GPa.}
\begin{ruledtabular}
\begin{tabular}{cc@{\hspace{0.5cm}}c@{\hspace{0.4cm}}c@{\hspace{0.4cm}}c@{\hspace{0.5cm}}c}
 \multicolumn{6}{c}{$a=b= 4.4813(4)$\,\r A,\quad $c=3.1062(2)$\,\r A} \\
 \multicolumn{6}{c}{$V=62.460(9) $\,\r A$^3$} \\
 \multicolumn{6}{c}{$P4_2/mnm$ (No. 136)} \\
 \multicolumn{6}{c}{$\lambda=0.2910$\,\r A} \\
 \multicolumn{6}{c}{$\theta_{\min}=2.6837^{\circ}$,\quad $\theta_{\max}=16.2454^{\circ}$} \\
 \multicolumn{6}{c}{$-6\leq h\leq 6$,\quad $-5\leq k\leq 5$,\quad $-4\leq l\leq 4$} \\
 \multicolumn{6}{c}{Number of total/unique reflections = $207 / 72$} \\
 \multicolumn{6}{c}{$R_{I>3\sigma(I)}= 0.014$, \quad $wR_{I>3\sigma(I)}= 0.019 $} \smallskip\\\hline
 & & $x/a$ & $y/b$ & $z/c$ & $U_{\rm iso}$ (\r A$^2$) \\
Ru & $2a$ &0  & 0   &  0   &  0.0033(1)\\
O & $4f$ & 0.3051(5) & 0.3051(5) & 0  &  0.0047(4)
\end{tabular}
\end{ruledtabular}
\label{1GPa}
\end{table}

\begin{table}[h!]
\centering
\caption{Details of the data collection and refined structural parameters for HP-I RuO$_2$ at 36.5 GPa.}
\begin{ruledtabular}
\begin{tabular}{cc@{\hspace{0.5cm}}c@{\hspace{0.4cm}}c@{\hspace{0.4cm}}c@{\hspace{0.5cm}}c}
 \multicolumn{6}{c}{$a=3.0477(8)$\,\r A,\quad $b=3.867(4)$\,\r A,\quad $c=4.6427(16)$\,\r A} \\
 \multicolumn{6}{c}{$V=54.72(6) $\,\r A$^3$} \\
 \multicolumn{6}{c}{$Pnnm$ (No. 58)} \\
 \multicolumn{6}{c}{$\lambda=0.2910$\,\r A} \\
 \multicolumn{6}{c}{$\theta_{min}=2.7364^{\circ}$,\quad $\theta_{\max}=16.2731^{\circ}$} \\
 \multicolumn{6}{c}{$-3\leq h\leq 4$,\quad $-6\leq k\leq 7$,\quad $-4\leq l\leq 4$} \\
 \multicolumn{6}{c}{Number of total/unique reflections = $174 / 91$} \\
 \multicolumn{6}{c}{$R_{I>3\sigma(I)}= 0.041$, \quad $wR_{I>3\sigma(I)}= 0.048 $}\smallskip\\\hline
 & & $x/a$ & $y/b$ & $z/c$ & $U_{\rm iso}$ (\r A$^2$) \\
Ru & $2a$ & 0  & 0   &  0   & 0.0052(4)\\
O & $4g$ & 0.266(4) & $-0.343(2)$ & 0  & 0.008(1)
\end{tabular}
\end{ruledtabular}
\label{36GPa}
\end{table}

\newpage

\bibliography{RuO2-pressure}

\end{document}